**Prebiotic Chemistry Insights for *Dragonfly*: Thermodynamics of Amino Acid Synthesis in Selk Crater on Titan**

Ishaan Madan[1][*], Ben K.D. Pearce[1]

[1] Department of Earth, Atmospheric, and Planetary Sciences, Purdue University, West Lafayette, IN 47907, USA; madani@purdue.edu, pearce21@purdue.edu

*Corresponding Author

**Abstract**

Saturn's moon Titan presents a compelling testbed for probing prebiotic chemistry beyond early Earth. Impact-generated melt pools provide transient aqueous habitats in an otherwise cryogenic environment. We use Cantera equilibrium models to assess whether mixtures of hydrogen cyanide (HCN), acetylene ($C_2H_2$), and ammonia ($NH_3$) can drive amino acid synthesis in Selk-sized craters. Across twenty-one amino acids (twenty proteinogenic plus β-alanine), $NH_3$-free systems yield only proline, alanine, and β-alanine, whereas adding as little as 1% $NH_3$ (relative to $H_2O$) renders almost the full suite accessible, with yields peaking at 2% and tapering thereafter. The $NH_3$-free alanine result implies alternative pathways beyond classical Strecker or aminonitrile hydrolysis, suggesting acetylene, abundant on Titan but scarce on early Earth, as a plausible feedstock. We identify acrylonitrile (detected on Titan) as a thermodynamically favorable intermediate that can convert to alanine under aqueous conditions in an $NH_3$-free pathway. For glycine and alanine production from nitrile hydrolysis, comparison with laboratory kinetics shows that our equilibrium models predict near-complete conversion, while observed rates yield only partial products over weeks. Yet estimated chemical equilibration times (years–centuries) are far shorter than melt lifetimes ($\geq 10^3$–$10^4$ yr), supporting plausibility of equilibrium in situ. These predictions are directly testable with *Dragonfly's* mass spectrometer (DraMS), for which we recommend pre-flight standards to test proline, alanine, β-alanine, cysteine, and methionine. The first three offer the best chances for amino acid detection regardless of ammonia availability; the latter two offer diagnostic tools for determining the presence of reactive sulfur in post-impact Titan ponds.

## 1. Introduction

Saturn's largest moon, Titan, is a world of astrobiological potential. It stands apart as the only other body in our Solar System besides Earth to host a dense atmosphere (~1.5 bar) and stable surface liquids. But unlike Earth, Titan's low surface temperature of 89–94 K (Cottini et al., 2012; Jennings et al., 2019) supports a methane-based hydrologic cycle akin to the water cycle on Earth: methane forms clouds, rain, and fluvial channels that drain into expansive lakes and seas near the poles (e.g., (Barnes et al., 2013; Brown et al., 2008; Fulchignoni et al., 2005; Lopes et al., 2007; Stofan et al., 2007; Turtle, Del Genio, et al., 2011; Turtle, Perry, et al., 2011; Turtle et al., 2009)). In the upper atmosphere, photochemistry driven by solar UV photons and energetic particles from Saturn's magnetosphere dissociates a few percent of methane (~2–5% $CH_4$) and nitrogen (~95–98% $N_2$) (Vuitton, 2025), producing a diverse suite of complex organic molecules (Hörst, 2017; Lora et al., 2025). These organics either remain suspended or condense into refractory aerosols (i.e. larger aggregates), forming the thick global haze that gives Titan its orange hue and shrouds the surface from remote observations (Hörst, 2017). Over geological time, sedimentation of this photochemical haze has created an organic-rich layer on the surface that blankets a water-ice crust (Coustenis, 1997; Griffith et al., 2003; Janssen et al., 2016). Cassini observations and atmospheric models suggest Titan's chemical inventory may include nitriles (molecules containing a carbon-nitrogen triple bond, e.g., hydrogen cyanide), alkynes (carbon-carbon triple bonds, e.g., acetylene), alkenes (carbon-carbon double bonds, e.g., ethylene) and other species relevant to prebiotic chemistry (e.g., (Brassé et al., 2017; Cable et al., 2012; Catherine D. Neish et al., 2010; H. Niemann et al., 2005; H. B. Niemann et al., 2010; Poch et al., 2012; Raulin & Owen, 2003; Anezina Solomonidou et al., 2025; Vuitton et al., 2025; Waite et al., 2007)). Collectively referred to as tholins, these photochemical organics can also yield biologically relevant molecules, including the building blocks of cellular life such as amino acids and nucleobases (e.g., (Brassé et al., 2017; Cable et al., 2012; Cleaves et al., 2014; Khare et al., 1986; Catherine D. Neish et al., 2009, 2010; Pearce, Hörst, Cline, et al., 2024)). While such atmospheric production and deposition is likely a major source of prebiotic feedstocks (organics that serve as starting materials for more complex chemistry), it may not be the only pathway to molecular complexity on Titan.

On Earth, life depends on liquid water, making aqueous environments a target for astrobiology. Although Titan's surface is too cold for persistent liquid water, impacts and cryovolcanism can melt the icy crust, creating transient water-rich habitats (Meyer-Dombard et al., 2025; Catherine D. Neish et al., 2018). Impact-generated melt pools may remain liquid for thousands of years (Artemieva & Lunine, 2003; Thompson & Sagan, 1992) and recent modeling of Selk crater suggests post-impact liquids could persist for tens of thousands of years (Wakita 脇田 et al., 2023). (Hedgepeth et al., 2022) show that the upper strata most accessible to *Dragonfly* even after erosion would solidify in much shorter timescales (~0.5–600 yr, depending on melt depth), while deeper melt can persist beneath an insulating ice shell for up to several kyr (Kalousová et al., 2024). In all cases, photochemically derived organics could mix with liquid water prior to refreezing (e.g., (Catherine D. Neish et al., 2018)), enabling aqueous processing that extends molecular complexity beyond atmospheric chemistry alone.

Titan's potential for biomolecule production has been studied primarily through the lens of atmospheric chemistry and deposition of organics on the surface. These tholins have been examined for their composition, reactivity, and relevance to Titan's prebiotic inventory through simulations and laboratory experiments as cited earlier. Far fewer studies address in situ synthesis within transient aqueous environments, such as impact melt pools or cryovolcanic flows. In the atmosphere, photochemical species adsorb onto complex organic haze particles and sediment onto the surface (Perrin et al., 2025). Once delivered, this material may undergo aqueous processing to form amino acids, nucleobases, or lipid precursors. The potential for such surface-driven synthesis remains an open question. NASA's upcoming *Dragonfly* mission, a rotorcraft lander scheduled to arrive in the mid-2030s, will explore Titan's low latitude regions, including Selk crater and surrounding regions (Barnes et al., 2021). Central to *Dragonfly's* astrobiology goals is the Dragonfly Mass Spectrometer (DraMS), a linear ion trap instrument (Grubisic et al., 2021; Moulay et al., 2023). DraMS will investigate Titan's surface composition, searching for chemical signatures of prebiotic pathways and assessing whether aqueous surface processes may yield biologically relevant compounds (Grubisic et al., 2021).

In this study, we evaluate the thermodynamic feasibility of synthesizing twenty-one canonical amino acids (all twenty proteinogenic and β-alanine) under post-impact conditions relevant to Selk-sized craters on Titan. We model aqueous mixtures of hydrogen cyanide (HCN) and acetylene ($C_2H_2$) interacting with varying concentrations of ammonia ($NH_3$) in liquid freshwater melt pools (i.e., without added salts or dissolved minerals). Using Gibbs free energy minimization, we identify which amino acids are thermodynamically favored under Titan-relevant conditions and assess the influence of ammonia on reaction outcomes. This represents the first thermodynamic equilibrium analysis of amino acid synthesis directly from simple surface organics in Titan melt pools.

## 2. Methods and Validation

### 2.1. Cantera Models

Thermodynamic modeling provides a principled way to evaluate the chemical behavior of a system at equilibrium by minimizing its Gibbs free energy (G). Gibbs energy is a measure of the system's potential to do non-mechanical work, combining the energy content and entropy effects. In simpler terms, Gibbs energy tells us whether a process can happen spontaneously, balancing stored energy with entropy. Such models can be used to build our groundwork understanding of organic synthesis in aqueous environments where in situ chemical analysis is limited, not yet explored, or not entirely possible (e.g., early earth history, subsurface oceans on other worlds like Europa and Enceladus). In principle, at a specified temperature and pressure, the chemical equilibrium state corresponds to the composition where the total G of the system reaches a minimum. This can be understood intuitively as we are used to the concept of a system tending towards its ground state energy. This framework enables the prediction of product distributions independent of reaction pathways, offering an upper bound estimate of product yield. For the aqueous organic chemistry explored in this study, such equilibrium calculations provide insight into which amino acids are thermodynamically accessible under Titan-relevant conditions.

To compute thermodynamic equilibrium states, we use Cantera (v3.1.0) (Goodwin et al., 2024): a Python-accessible chemical kinetics and thermodynamics software package. Specifically, we use Cantera's 'VCS' chemical equilibrium solver, which uses a stoichiometric matrix formalism grounded in the principles of Chemical Reaction Stoichiometry (CRS) as described in (Missen & Smith, 1998). Conceptually, the algorithm is asking, "Given a fixed number of atoms, what is the most energetically and entropically favorable way to rearrange them into molecules?" It solves this by balancing the "building blocks" (reactants) and choosing the combination of species that results in the lowest overall Gibbs energy while conserving mass and charge. Detailed mathematical formulation of the solver is presented in Appendix Section A. Each reaction system is defined by a specific set of initial reactants and concentrations and modeled independently. The method is time-independent, meaning it does not simulate reaction kinetics but instead identifies the final equilibrium state that would be reached if the system were allowed to react indefinitely. This means the computed yields represent the maximum thermodynamic extent of reaction, independent of temperature-dependent rates or pathways. In practice, this captures the end-state chemistry towards which Titan's aqueous systems may evolve, with real equilibration likely proceeding more slowly at lower temperatures.

This framework requires Gibbs energy data across a range of temperatures. The Gibbs free energies of formation, $\Delta_f G^o(T)$ used here are obtained from the CHNOSZ database (Dick, 2019) using the python wrapper, pyCHNOSZ (Boyer, 2024). For species lacking published Gibbs data, we estimate $\Delta_f G^o(T)$ building on a well established quantum-chemistry-based method following (Ochterski, 2000) (as described in Section 2.2). To interface with Cantera, we convert the $\Delta_f G^o(T)$ values into the NASA 9-coefficient polynomial format (McBride et al., 2002), the numerical details for which can be found in Appendix Section A.

We validate our Cantera models against the thermochemical study of (Pearce & Pudritz, 2016), who modeled aqueous nucleobase synthesis in asteroid interiors using proprietary software (Appendix Section A). Under matched pressure, temperature, and composition, our Cantera outputs closely replicate their product yields across four reactions from 0–200 ºC, diverging only near the 100-bar boiling point of water where known dataset differences emerge (Figure A1). This agreement confirms that our conversion and equilibrium framework yield consistent, robust predictions.

*2.2. Gibbs Free Energies Estimator*

Many prebiotically relevant molecules in Titan analog studies lack temperature-dependent Gibbs free energy data, particularly aminoacetonitrile (AAH), 2-aminopropanenitrile (2AH), and β-alanine in our case. To incorporate these into Cantera's equilibrium models, we develop a hybrid quantum chemistry-analog transfer method to estimate $\Delta_f G^o(T)$ across 273–598 K. At 298 K, we calculate $\Delta_f G^o(T)$ using the well-established 3-step protocol of (Ochterski, 2000), summarized as:

$$\Delta_f G^o(298K) = \Delta_f H^o(298K) - TS^o(298\,K) \qquad (1)$$

where $\Delta_f H^o$ is the enthalpy of formation derived from electronic energies (EE), zero-point energy (ZPE) corrections, and thermal corrections, $T$ is the temperature (298 K), and $S^o$ is total entropy (including the translational, rotational, and vibrational contributions). See Appendix Section B for the full 3-step protocol leading to Eq. 1. The thermochemical information (e.g., EE, ZPE) is obtained from quantum chemical calculations. All quantum chemical calculations used density functional theory (DFT) with the B3LYP hybrid functional (Becke, 3-parameter, Lee–Yang–Parr) (Becke, 1993; Lee et al., 1988) and the 6-311++G(2df, 2p) basis set, as implemented in Gaussian 16 Revision B.01 (Frisch et al., 2017). Implicit solvation is modeled using the polarizable continuum model (PCM) with water as the solvent (Cammi & Tomasi, 1995; Miertuš et al., 1981; Tomasi et al., 1999). This protocol, previously benchmarked for aqueous organics such as formaldehyde, reproduced experimental Gibbs energies within ~5–10 kJ mol⁻¹ at 298 K (Paschek et al., 2022). All calculated structures are confirmed as true minima via frequency analysis.

One current limitation with Gibbs free energy calculations at multiple temperatures using quantum chemistry is that they produce trends that are inconsistent with empirical datasets (monotonic increase in $G$ with $T$; see Figure A2). To correct for this, we extend the 298 K results using an analog-transfer method. For each target molecule, we select a structurally similar analog with full CHNOSZ $\Delta_f G^o(T)$ data: (i) AAH → glycine, (ii) 2AH → alanine, (iii) β-alanine → alanine, (iv) acrylonitrile → alanine, justified by direct precursor-product relationships and isomeric similarity. The analog's (glycine, alanine) $\Delta_f G^o(T)$ are first fit to a line to extract the slope, $m$. This slope is then combined with the target molecule's quantum chemistry value at 298 K to set the intercept, $b$, yielding a linear relation:

$$\Delta_f G^o(T) = mT + b; \text{ where } b = \Delta_f G^o(298\ K) - (m \cdot 298) \tag{2}$$

This resulting linear function is then used to estimate Gibbs free energies at all other temperatures. This preserves the exact 298 K quantum result while inheriting the empirical temperature dependence of the analog. The reconstructed curves are then fitted to the NASA-9 polynomials (as described in the previous section) for Cantera's compatibility. Benchmarking against CHNOSZ data shows deviations of <16% and <21%, for glycine and alanine respectively at 298 K and agreement within 14–17% (glycine) and 18–22% (alanine) across 273–593 K, supporting the method's validity (Figure A2). We apply this method to AAH, 2AH, β-alanine, and acrylonitrile. Table 1 summarizes the calculated values over 0–100 ºC, with the full range in Table A1.

Table 1. Calculated standard Gibbs free energies of formation, $\Delta_f G^o$ (J mol⁻¹) for aminoacetonitrile (AAH), 2-aminopropanenitrile (2AH), β-alanine, and acrylonitrile as a function of temperature.

| T (ºC) | $\Delta_f G^o$(J mol⁻¹) | | | |
|---|---|---|---|---|
| | AAH | 2AH | β-alanine | Acrylonitrile |
| 0.01 | 141727 | 154434 | -285049 | 179558 |
| 10 | 139879 | 152206 | -287277 | 177330 |
| 20 | 138029 | 149976 | -289507 | 175100 |
| 30 | 136179 | 147746 | -291737 | 172870 |
| 40 | 134329 | 145516 | -293967 | 170640 |
| 50 | 132479 | 143286 | -296197 | 168410 |
| 60 | 130629 | 141056 | -298427 | 166180 |
| 70 | 128779 | 138826 | -300658 | 163950 |
| 80 | 126929 | 136596 | -302888 | 161720 |
| 90 | 125079 | 134366 | -305118 | 159490 |
| 100 | 123229 | 132136 | -307348 | 157260 |

We compare our quantum chemistry (QC) anchors at 298 K to estimates from the Joback group-contribution method. Joback computes thermodynamic properties by summing contributions from predefined functional groups, making it a fast and accessible estimation tool (Joback, 1984; Joback & Reid, 1987). In the case of glycine and alanine, Joback differs from QC by ~2 kJ mol⁻¹ and ~12 kJ mol⁻¹. In other words, Joback yields slightly closer agreement to CHNOSZ than QC at 298 K.

However, Joback carries several limitations. The method's reference state is the ideal gas at 298 K, so solvation effects (e.g., hydrogen bonding, zwitterionic stabilization) are not represented. The scheme is group-additive and geometry-agnostic; therefore, it has limited ability to capture isomeric differences beyond group counts and does not distinguish between stereoisomers (e.g., cis/trans). In practice, its implementation requires subjective choices in how functional groups are assigned, leading to inconsistent values. For instance, using the online EGIChem calculator, glycine's formation energy is reported as ~-311.8 kJ mol$^{-1}$, whereas Cheméo reports -233 kJ mol$^{-1}$, both citing the Joback method. Such variation shows that Joback is not a universal or robust framework. Finally, it is unsuitable as a primary source for chemical equilibrium constants, a concern noted by the authors themselves (Joback & Reid, 1987). For these reasons, we adopt QC-based 298 K anchors as our primary method. While QC has its own approximations, it incorporates some solvation effects, accounts for the full molecular structure and energy differences across isomers, and provides a transferable, first-principles basis to compute thermodynamic properties.

While our approach simplifies by assuming linear $\Delta_f G^o(T)$ behavior and transferable slopes between analogs, it offers a defensible method for extending equilibrium thermodynamic modeling to molecular species lacking data. This is valuable in prebiotic chemistry, where many intermediates of interest have no empirical thermodynamic data and can be difficult to work with experimentally. Our method enables sensitivity testing of such species and represents one of the few practical avenues to incorporate novel molecular species into thermodynamic modeling of planetary environments. The code developed for calculating Gibbs free energies at 298 K is provided open source on Zenodo at the DOI provided in the Data Availability Statement. It digitizes and automates the (Ochterski, 2000) three-step protocol using quantum chemistry log files (e.g., Gaussian outputs).

## 2.3. Why Thermodynamics for Titan?

To benchmark our thermodynamic modeling approach, we simulate two kinetic reactions experimentally investigated in (Farnsworth et al., 2024). This is the only published study that quantifies amino acid production from aminonitriles under Titan-relevant aqueous conditions. We note that other works (e.g., (Cleaves et al., 2014; Catherine D. Neish et al., 2010)) have performed amino acid synthesis via hydrolysis of tholin analogs produced in the lab. We expect tholins would contribute some amino acids to Titan post-impact ponds and we aim to calculate the amino acid contribution from tholins separately in a future time-dependent model. (Farnsworth et al., 2024) experimentally simulates the alkaline hydrolysis of aminonitriles, a reaction representing a late step in a multistep sequence that may start from simpler feedstocks like HCN. Although our fiducial models focus on simple building blocks such as HCN and NH$_3$, we intentionally replicate the reactions from (Farnsworth et al., 2024) to assess how thermodynamics compare with experimentally observed kinetic behavior. We model two of the three reactions examined in (Farnsworth et al., 2024), namely aminoacetonitrile (AAH) hydrolyzing to glycine, and 2-aminopropanenitrile (2AH) hydrolyzing to alanine, because these products are a focus of our study. For each reaction, we use the reported reactant concentrations and ammonia levels (0%, 5%, 10%, 15%) and compute thermodynamic equilibrium at the temperatures tested experimentally (-22 to 25 ºC). Notably, only one modeled case corresponds to the lowest experimental temperature of -22 °C condition, which is a 15% $NH_3$ mixture. At first glance, one might expect this system to freeze, since the temperature is well below the freezing point of pure water. However, (Farnsworth et al., 2024) observed that the mixture remained liquid, owing to the cryoscopic effect of ammonia. This is consistent with (Postma, 1920), who reported that a 17% aqueous ammonia solution freezes near -28 °C. This freezing-point depression supports the validity of our equilibrium results for this case. Table 2 summarizes the modeled inputs. Aminoacetonitrile (AAH) and 2-aminopropanenitrile (2AH) do not have Gibbs free energies reported across temperature in existing literature. We therefore calculated $\Delta_f G^o(T)$ for each using our quantum chemistry-based estimator (Table 1).

Table 2. Modeled input concentrations for AAH and 2AH hydrolysis as mole ratios relative to $H_2O$.

| Reaction | Molecule | Input concentration (mol X/mol $H_2O$) | | | |
|---|---|---|---|---|---|
| | | *0% NH₃* | *5% NH₃* | *10% NH₃* | *15% NH₃* |
| AAH → Glycine | $H_2O$ | 1.000 | 1.000 | 1.000 | 1.000 |
| | $NH_3$ | 0 | 0.056 | 0.118 | 0.187 |
| | AAH | 0.002 | 0.002 | 0.002 | 0.002 |
| 2AH → Alanine | $H_2O$ | 1.000 | 1.000 | 1.000 | 1.000 |
| | $NH_3$ | 0 | 0.056 | 0.118 | 0.187 |
| | 2AH | 0.001 | 0.001 | 0.001 | 0.001 |

### 2.4. Thermodynamic vs. Kinetic Outcomes

Our Cantera model computes equilibrium compositions by minimizing Gibbs free energy, assuming ideal activity, infinite time, and no kinetic barriers. In every $NH_3$-bearing case, this yields >99% conversion of aminonitrile to the corresponding amino acid, independent of temperature. In contrast, $NH_3$-free systems show no conversion, matching Farnsworth et al.'s (2024) first major conclusion that ammonia is crucial for these hydrolysis reactions.

Divergences between our results and (Farnsworth et al., 2024) yields stems from fundamental differences between thermodynamic and kinetic control. Thermodynamic equilibrium reflects the most stable, lowest-energy system, independent of any reaction pathway. Kinetic outcomes, on the other hand, are governed by activation energy barriers, intermediate formation, finite timescales, and competing reactions. Consequently, the lab experiments (Farnsworth et al., 2024) yield only partial conversion across days to weeks, while our models estimate the maximum achievable conversion given sufficient time. For example, (Farnsworth et al., 2024) report final glycine and alanine yields of ~3–6% and ~35–40% respectively, under $NH_3$-bearing conditions depending on temperature and ammonia content (after 6 months), far below our >99% equilibrium predictions. Two key conclusions from our benchmark qualitatively support and extend Farnsworth et al.'s (2024) findings:

- Ammonia is required for conversion to glycine and alanine in aminonitrile pathways, both experimentally and thermodynamically.
- Increased ammonia beyond 5% yields minimal additional gain in amino acid production. This is temperature-independent in our models but observed by (Farnsworth et al., 2024) specifically at 3ºC.

Our models represent the thermodynamic ceiling of amino acid production in the studied reactions. Which raises the question: would Titan's impact-generated melt pools remain liquid long enough to reach equilibrium? Past work (Artemieva & Lunine, 2003; Thompson & Sagan, 1992) and recent models of Selk crater formation (Wakita 脇田 et al., 2023) suggest that post-impact melts on Titan may persist for several thousands of years. However, (Hedgepeth et al., 2022) demonstrate that near-surface melt, particularly the upper 10–50 m accessible to *Dragonfly,* would solidify in ~0.5–10 yr. Although this is much shorter than the total lifetime of the deeper melt lens, it still provides a meaningful window for aqueous reactions, especially considering that the liquid temperature may be >20 º C rather than the conservative 0º C assumed in (Hedgepeth et al., 2022). But is this sufficient time to reach chemical equilibrium on Titan? To approach this, we use the experimentally determined rate constants from (Farnsworth et al., 2024), derived from first-order fits to product yield curves, along with our model's 99% conversion ceiling, to estimate equilibration timescales. These values provide a numerical estimate of how long such chemical reactions might take to reach equilibrium under Titan-relevant conditions. Table 3 summarizes the results. Appendix Section C shows further numerical details.

Table 3. Experimental rate constants from (Farnsworth et al., 2024) and estimated equilibration times for aminonitrile hydrolysis.

| Reaction[1] | NH$_3$ (%)[1] | T (ºC)[1] | k (s$^{-1}$)[1] | Equilibration Time (yr)* |
|---|---|---|---|---|
| AAH → Glycine | 5 | 3 | 1.68E-09 ± 3.10E-10 | 73–107 |
| | 10 | 3 | 4.05E-09 ± 1.60E-10 | 35–38 |
| | 15 | 21 | 5.82E-08 ± 2.51E-08 | 2–4 |
| 2AH → Alanine | 5 | 3 | 3.18E-08 ± 2.70E-09 | 4–5 |
| | 10 | 3 | 2.65E-08 ± 9.00E-10 | 5–6 |
| | 15 | -22 | 1.78E-09 ± 5.90E-10 | 62–123 |
| | 15 | 21 | 1.13E-07 ± 1.10E-08 | 1.2–1.4 |

[1](Farnsworth et al., 2024). *The equilibration time is a range to include the uncertainties associated with the rate constant, $k$.

Considering the uncertainties and the longest equilibration time in Table 3 (123 years), equilibration appears plausible within the overall melt, particularly within the deeper layers that remain liquid for longer durations but is less likely to be observed directly by *Dragonfly*. The upper tens of meters, however, could host partial equilibration under lower temperatures, and may even approach thermodynamic equilibrium if subjected to elevated post-impact temperatures. We acknowledge that this estimation is limited to the aminonitrile hydrolysis pathway and does not capture the full landscape of Titan's potential prebiotic chemistry. Nonetheless, thermodynamic modeling provides an upper bound for chemical yield and complements kinetic studies by extending predictive power into conditions not yet accessible via experiment or in situ measurements.

## 2.5. Applying the model to Titan: Melt Volume and Input Organics

Numerical modeling of Selk crater suggests that the impact of a 4 km icy body into a clathrate-rich icy surface generates about 200 km³ of complete melt and about 400 km³ of partial melt, which may remain liquid for thousands of years under post-impact conditions (Wakita 脇田 et al., 2023). Assuming most prebiotic chemistry is occurring in the complete melt, we adopt the lower bound volume of 200 km³. Assuming pure water density, this corresponds to ~ 1.11 x 10¹⁶ mol $H_2O$, which serves as the normalization basis for all other species.

As for the surface organic inventory, quantitative constraints are limited. As a starting point, we draw from Table 4 of (C. Neish et al., 2024), which estimates the long-term deposition of photochemically derived organics on Titan's surface based on atmospheric models (Krasnopolsky, 2009). We exclude haze related fractions ($C_xH_yN$, $C_xH_y$, and charged haze) due to their poorly defined molecular identities and lack of available thermodynamic data. We focus on $C_2H_2$ and HCN, which are the most abundant and tractable species in the inventory. Other species listed in that table are omitted because (i) their fluxes are at least 4 times lower than $C_2H_2$ and HCN, and (ii) our goal is to define a minimum viable feedstock for amino acid synthesis to maintain conservative estimates.

The total moles of each species, $n_i$, delivered to Selk's 45-km-radius catchment area over time, $t$, are calculated using:

$$n_i = \frac{m_i}{M_i} = \frac{F_i \cdot A \cdot t}{M_i} \tag{3}$$

Where $i$ denotes a species, $m_i$ is the total mass delivered (g), $M_i$ is the molar mass (g mol⁻¹), $F_i$ is the flux from Table 4 of (C. Neish et al., 2024) (g cm⁻² Gyr⁻¹), $A$ is the circular catchment area with a 45 km radius, and $t$ is the accumulation time (4.3 Gyr). Selk crater's age is not precisely constrained, but its morphology and degree of modification indicate it is relatively young compared to other Titan craters. Titan's overall low crater density implies a young crater retention age of ~0.2-1.0 Gyr (Hedgepeth et al., 2020; C. D. Neish & Lorenz, 2012), with recent work suggesting closer to 0.3 Gyr (Wakita et al., 2024). Selk itself is among Titan's larger (~90 km diameter) and less degraded craters, with a depth-to-diameter ratio implying moderate erosion but preservation of much of its original structure (Hedgepeth et al., 2020; Lorenz et al., 2021). Cassini's Visual and Infrared Mapping Spectrometer (VIMS) data analysis demonstrated that Selk's ejecta blanket exhibits signatures of organic-rich material together with localized water-ice enrichment, supporting its classification as an intermediately degraded and compositionally heterogenous crater (A. Solomonidou et al., 2020; Werynski et al., 2019). It is young enough that its morphology is still recognizable, yet old enough that some infilling and erosion have reduced its depth. This places it within the broad range of tens to hundreds of Myr, consistent with estimates that Titan's observed craters are no older than several hundred Myr (Hedgepeth et al., 2022, 2020; C. D. Neish et al., 2015, 2016; Werynski et al., 2019). For our purposes, we adopt 200 Myr as age and assume steady organic deposition over Titan's 4.3 Gyr history prior to the impact event. Several processes (e.g. photolysis, burial, oxidation) may destroy a substantial fraction of any settled organics before and/or post-impact; therefore, we apply a 10% survival factor to HCN and $C_2H_2$ inventory, thereby ensuring that our models remain conservative. This factor, albeit somewhat arbitrary, is consistent with recent Titan impact studies that show a total survival fraction for amino acids of ~12–33% for average impact conditions on Titan (Pearce, Hörst, Cline, et al., 2024). To verify that our conclusions are not sensitive to this assumption, we modeled 1% and 30% survival factor scenarios (Appendix Section E, Figure A3, Tables A3 and A4). These tests showed consistent results for most amino acids and ammonia concentrations. For the 1% survival factor tests, we see slight increases in yields for the 1% $NH_3$ tests for isoleucine, leucine, and valine. For the 30% survival factor tests, we see lower yields for these species as well as arginine and lysine for the 1–3 % $NH_3$ scenarios. Besides these slight variations, the overall qualitative patterns of our results remain the same.

 Ammonia ($NH_3$) is also included in our models due to its expected presence in Titan's subsurface (Gabriel Tobie et al., 2005) and relevance in abiotic amino acid chemistry (e.g., Strecker Synthesis (Bada, 2023)). Ammonia may partition into impact melts either via direct release from Titan's interior or release from clathrates disrupted during impact (e.g. (Choukroun & Sotin, 2012; Kalousová & Sotin, 2020)). Recent experimental work has shown that ammonia also plays a crucial role in amino acid synthesis from aminonitriles under Titan-like conditions (Farnsworth et al., 2024). However, the abundance and distribution of ammonia within Titan's crust and subsurface ocean remain poorly constrained, and the degree to which it partitions into the impact melts is not fully understood. To account for this uncertainty, we treat $NH_3$ as a variable and explore a range of initial concentrations (0%, 1%, 2%, 3%, 4%, 5%, and 10% relative to water). Final reactant concentrations are normalized to the total $H_2O$ melt pool and summarized in Table 4. Other numerical details are provided in Table A3 of Appendix Section D.

Table 4. Initial concentrations of the reactants used in our fiducial thermochemical models.

| Molecule | Concentration (mol X/mol $H_2O$) |
|----------|-----------------------------------|
| $H_2O$ | 1.00 |
| $C_2H_2$ | 0.042 |
| HCN | 0.020 |
| $NH_3$ | 0, 0.01, 0.02, 0.03, 0.04, 0.05, 0.10 |

## 3. Results and Discussion

### 3.1. Results: Amino Acid Formation

Percent yields for the twenty-one canonical amino acids, relative to the limiting reagent HCN, at 0 ºC are summarized in Figure 1 and Table 5. Our fiducial models are run at 0 ºC; however, equilibrium yields are effectively temperature-independent within the 0–100 ºC range, as variations in Gibbs free energies of formation across this interval (~6–32 kJ mol$^{-1}$) are not substantial relative to their absolute magnitudes to appreciably alter yields. Thus, our results are applicable to aqueous environments spanning 0–100 ºC. Seven starting $NH_3$ conditions are examined (Table 4). At 0% $NH_3$, measurable formation is confined to a small subset of amino acids: proline, alanine, and β-alanine. Introducing as little as 1% $NH_3$ makes a notable difference, leading to the synthesis of nearly all studied amino acids except sulfur-bearing cysteine and methionine. Overall yields increase up to 2% $NH_3$, beyond which they taper, suggesting an optimal ammonia range for maximum yield (Table 5).

Figure 1. Heat map of amino acid percent yields at 0 ºC relative to the starting HCN inventory in our fiducial models with a 10% survival factor. Columns correspond to varying $NH_3$ concentrations (0–10% of the total water content). Rows list the 21 amino acids evaluated. Cell shading follows a viridis scale from 0% (purple) to 100% (yellow).

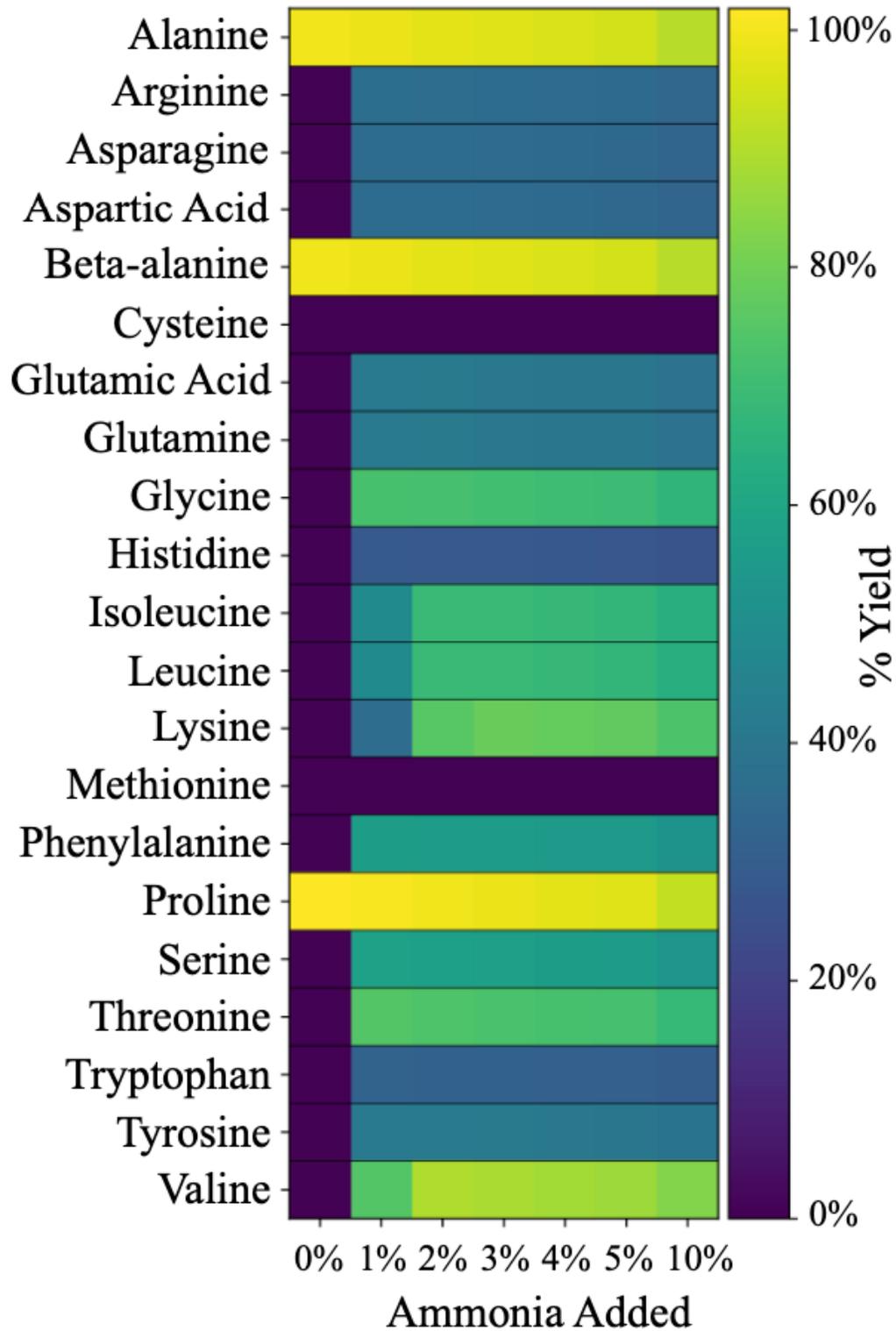

Table 5. Percent yields of amino acids at 0 ºC, expressed relative to the initial HCN inventory.

| Amino Acid | % Yield (with respect to HCN) | | | | | | |
|---|---|---|---|---|---|---|---|
| | *0% NH₃* | *1% NH₃* | *2% NH₃* | *3% NH₃* | *4% NH₃* | *5% NH₃* | *10% NH₃* |
| Alanine | 99.8 | 98.8 | 97.8 | 96.9 | 96.0 | 95.1 | 90.7 |
| Arginine | 0 | 36.7 | 36.3 | 36.0 | 35.6 | 35.3 | 33.7 |
| Asparagine | 0 | 36.2 | 35.8 | 35.5 | 35.1 | 34.8 | 33.3 |
| Aspartic acid | 0 | 36.2 | 35.8 | 35.5 | 35.1 | 34.8 | 33.3 |
| β-alanine | 99.8 | 98.8 | 97.8 | 96.9 | 96.0 | 95.1 | 90.7 |
| Cysteine | 0 | 0 | 0 | 0 | 0 | 0 | 0 |
| Glutamic acid | 0 | 41.8 | 41.4 | 41.0 | 40.6 | 40.2 | 38.4 |
| Glutamine | 0 | 41.8 | 41.4 | 41.0 | 40.6 | 40.2 | 38.4 |
| Glycine | 0 | 72.3 | 71.6 | 71.0 | 70.3 | 69.6 | 66.5 |
| Histidine | 0 | 28.8 | 28.6 | 28.3 | 28.0 | 27.8 | 26.5 |
| Isoleucine | 0 | 48.9 | 69.2 | 68.5 | 67.8 | 67.2 | 64.1 |
| Leucine | 0 | 48.9 | 69.2 | 68.5 | 67.8 | 67.2 | 64.1 |
| Lysine | 0 | 36.4 | 75.2 | 78.5 | 77.8 | 77 | 73.4 |
| Methionine | 0 | 0 | 0 | 0 | 0 | 0 | 0 |
| Phenylalanine | 0 | 56.9 | 56.3 | 55.8 | 55.2 | 54.7 | 52.2 |
| Proline | 101.8 | 100.8 | 99.8 | 98.8 | 97.8 | 96.9 | 92.4 |
| Serine | 0 | 58.4 | 57.8 | 57.3 | 56.7 | 56.2 | 53.7 |
| Threonine | 0 | 74.5 | 73.7 | 73.0 | 72.3 | 71.6 | 68.4 |
| Tryptophan | 0 | 32.5 | 32.2 | 31.9 | 31.6 | 31.3 | 29.9 |
| Tyrosine | 0 | 42.1 | 41.7 | 41.3 | 40.9 | 40.5 | 38.7 |
| Valine | 0 | 74.5 | 89.6 | 88.8 | 87.9 | 87 | 83 |

The 0 °C simulations, although near the freezing point of pure water, are physically relevant for Titan-like aqueous systems. The VCS Gibbs energy minimization routine is temperature-explicit (i.e., equilibrium is calculated separately at each temperature) and time-independent, so equilibrium compositions are unaffected by kinetic barriers even at subzero temperatures. Secondly, the modeled mixtures contain solute loads (HCN, NH₃, and intermediate species) that can depress the freezing point; for example, a 5% aqueous ammonia solution freezes near −5 °C, and higher concentrations lower it further (Postma, 1920).

### 3.2. Formation of amino acids in 0% NH₃ system

One of the most unexpected outcomes of our equilibrium modeling is the formation of alanine, β-alanine, and proline in the 0% NH₃ simulations. This likely reflects the combined influence of stoichiometric balance and Gibbs energetics within the chosen reactant inventory (Table 4). Nitrogen can be sourced directly from HCN, and its combination with acetylene favors $C_3/C_5$ skeletons, stabilizing alanine, β-alanine, and proline. These two factors demonstrate how the starting inventory constrains which amino acids are thermodynamically accessible. Our 0% NH₃ results may represent a post-impact melt pond with sustained high temperatures sufficient to volatilize and remove ammonia from the system prior to equilibration.

Focusing on alanine, the fiducial models contrast both laboratory work and our kinetic validation models (See Section 2.4 above). (Farnsworth et al., 2024) observed zero alanine yield from 2-aminopropanenitrile (2AH) hydrolysis without ammonia, a result our validation model reproduces (Reaction 1). We perform a series of equilibrium tests, summarized in Table 6, that suggest alanine in our fiducial models does not form via classic aminonitrile hydrolysis (or Strecker-type) pathways (Reaction 2). Including NH₃ in the species list but setting its initial concentration to zero also produces alanine without net NH₃ forming at equilibrium, confirming that ammonia is neither required nor activated as an intermediate in our model (Reaction 3). Removing $C_2H_2$ from the starting mixture eliminates alanine production entirely, indicating that acetylene is an essential carbon feedstock (Reactions 4 and 5).

In a limited effort to probe NH₃-free pathways to alanine, we test two candidate intermediates. Ethanamine (ethylamine) is explored due to its structural similarity to alanine. Cosmochemically, ethanamine has been identified in the Murchison meteorite as the decarboxylation (i.e., alanine → ethanamine via removal of –COOH group; usually through heat) product of alanine (D. P. Glavin & Bada, 2001; Martins & Sephton, 2009), and both alanine and ethanamine have been detected together in Ryugu samples (Naraoka et al., 2023; Sillerud, 2024). Despite these motivations, our equilibrium tests find no alanine production from ethanamine and water (Reaction 6), and even ethanamine, acetylene, and water did not produce alanine (Reaction 7). These negative results suggest that ethanamine is unlikely to be the missing intermediate driving NH₃-free alanine formation in our fiducial models. We also explore acrylonitrile, another molecule structurally related to alanine and directly detected on Titan (Palmer et al., 2017; Thelen et al., 2019). Acrylonitrile has previously drawn interest as a potential membrane-forming molecule (azotosomes; (Stevenson et al., 2015)), though a subsequent study showed such structures are thermodynamically unfavorable on Titan (Sandström & Rahm, 2020). In our models, acrylonitrile is favorably produced from the fiducial inventory (Reaction 8) and in turn forms alanine via reaction with water (Reaction 9). This suggests acrylonitrile could plausibly act as an intermediate in the ammonia-free route to alanine. However, we emphasize that thermodynamics alone cannot resolve whether this route is kinetically viable on the aqueous timescales of post-impact ponds. Laboratory experiments indicate that acrylonitrile requires ammonia to form β-alanine (Ford et al., 1947), not alanine, simply highlighting that thermodynamic predictions must be tested against kinetic constraints.

Taken together, these tests suggest that alanine formation from HCN, $C_2H_2$, and $H_2O$ is thermodynamically downhill, even without ammonia. Thermodynamics alone, however, says little about rates and in practice, these reactions may face substantial kinetic hurdles that prevent them from proceeding at appreciable rates without catalysts or energetic inputs. Conversely, Titan's environment may provide ways to mitigate such barriers (e.g., prolonged aqueous intervals, mineral-like surfaces, irradiation). Our results should therefore be interpreted as identifying *thermodynamically allowed* routes that broaden the prebiotic landscape, while highlighting the need for kinetic experiments to test their plausibility under Titan-like conditions.

Table 6. Thermochemical equilibrium tests for alanine formation under $NH_3$-free and $NH_3$-bearing conditions.

| Reaction Number | Reaction Tested | Alanine Formation | Observations |
|---|---|---|---|
| 1 | $2AH + H_2O$ (no $NH_3$) → Alanine | No | Matches (Farnsworth et al., 2024) results; $NH_3$ required for 2AH hydrolysis |
| 2 | $HCN + C_2H_2 + H_2O$ → Alanine | Yes | Main $NH_3$-free simulation; alanine forms without $NH_3$ |
| 3 | $HCN + C_2H_2 + H2O$ ($NH_3$ in species list, init. conc = 0) → Alanine | Yes | No $NH_3$ forms at equilibrium; $NH_3$ not a required intermediate |
| 4 | $HCN + H_2O$ → Alanine | No | $C_2H_2$ essential for alanine production |
| 5 | $HCN + H_2O + NH_3$ (5–10%) → Alanine | No | $NH_3$ alone with HCN not sufficient; $C_2H_2$ required |
| 6 | Ethanamine + $H_2O$ → Alanine | No | Ethanamine alone not sufficient |
| 7 | Ethanamine + $C_2H_2 + H_2O$ → Alanine | No | Ethanamine likely not the intermediate forming Alanine |
| 8 | $HCN + C_2H_2 + H_2O$ → Acrylonitrile | N/A | Acrylonitrile is favorably produced; Alanine is not included in this simulation |

| 9 | Acrylonitrile + $H_2O$ → Alanine | Yes | Acrylonitrile may act as an intermediate in the formation of Alanine |

Previous proposals for acetylene-based prebiotic chemistry have generally involved catalyzed or energy-driven reaction networks. On Titan, acetylene has been proposed to be a precursor to aromatics and amino-acid like products via catalytic cyclotrimerization to benzene and subsequent substitution reactions, processes that require mineral or metal catalysts to lower the energetic barrier (see (O. Abbas & Schulze-Makuch, 2002; S. H. Abbas & Schulze-Makuch, 2007). Laboratory studies under simulated hydrothermal conditions similarly demonstrate that acetylene, in combination with CO and $NH_3$, can yield amino acids and amides when catalyzed by transition metal sulfides (Seitz et al., 2024). Astrophysical models further indicate that acetylene polymerization contributes to carbon chain growth, polyynes, cyanopolyynes, and aromatic molecules in circumstellar and interstellar environments (Cernicharo, 2004; Pentsak et al., 2024). In contrast, our work identifies a simpler, potentially catalyst-free route that is thermodynamically favorable under aqueous Titan-like conditions. Characterizing the detailed reaction pathway lies beyond the scope of our study and will require future experimental and computational quantum chemistry work.

Observational evidence supports the plausibility of acetylene-driven prebiotic chemistry on Titan. (Singh et al., 2016) used Cassini VIMS data to identify solid acetylene deposits on Titan's equatorial surface, concentrated in dark dune fields such as Shangri La and Fensal–Aztlan/Quivira. These findings imply that acetylene is produced photochemically in Titan's atmosphere and accumulates in surface reservoirs, where it may be mobilized by weathering and transport. The co-location of acetylene-rich terrains with *Dragonfly's* planned landing region asserts the relevance of our models.

Finally, the ubiquity of amino acids in extraterrestrial materials highlights the relevance of these results. Carbonaceous chondrites such as Murchison contain a diverse amino acid inventory, with isotopic signatures consistent with abiotic synthesis (D. P. Glavin & Bada, 2001; Martins & Sephton, 2009). More recently, sample return missions from Ryugu and Bennu revealed glycine, alanine, and other amino acids, in some cases alongside related amines such as ethanamine, confirming that amino acid chemistry is not unique to Earth (Daniel P. Glavin et al., 2025; Naraoka et al., 2023; Sillerud, 2024). Although the specific roles of ammonia, hydrogen cyanide, or acetylene, and other molecules in these environments remain an open area of inquiry, thermodynamic models provide some insight. (Cobb et al., 2015) modeled Strecker-type synthesis in carbonaceous chondrite parent bodies and found good agreement between predicted relative amino acid abundances and those measured in the meteoritic record. The detection of amino acids across meteorites and asteroids suggests that multiple feedstock combinations may converge on similar end products; albeit, with different relative abundances. For example, although alanine is a dominant amino acid in the CR2 and CM2 meteorites, similar to our Titan models, proline is not in the top two most abundant proteinogenic amino acids in these meteorites, rather, it falls to 5th to 7th (Cobb & Pudritz, 2014). This divergence likely arises from environmental and compositional differences; aqueous alteration within meteorite parent bodies involves prolonged liquid water exposure at higher temperatures, whereas Titan's impact-generated melt pools represent transient cryogenic to aqueous systems rich in acetylene and other hydrocarbons. Acetylene is absent from the starting inventories of (Cobb et al., 2015) but highly relevant for Titan and this work, where it provides a key carbon source for the amino acid backbone. Another consideration is that the interior compositions of carbonaceous chondrites parent bodies have not been measured. (Cobb et al., 2015) used cometary coma interiors (acetylene content of ~0.1% relative to water) as a proxy for asteroid interiors.

Titan thus represents another compelling test case, as its surface organics and impact-induced aqueous environments provide a natural laboratory analogous to early Earth (e.g., (Trainer et al., 2006)) but with a distinct inventory of hydrocarbons. *Dragonfly* will directly probe these terrains with DraMS, offering the first opportunity to evaluate whether hydrocarbon-rich environments support amino acid synthesis.

### 3.3. Relevance to Dragonfly and other missions

DraMS, the mass spectrometer on-board, comprises two complementary modes: laser desorption/ionization mass spectrometry (LDMS) and gas chromatography–mass spectrometry (GC-MS), both of which are well-suited for amino acid detection. Our equilibrium models provide a predictive framework for DraMS science operations by identifying which amino acids may be likely to form in Selk crater melt pools. Specifically, our model yields testable predictions: (1) alanine, β-alanine, and proline are predicted to form even in the absence of $NH_3$, (2) presence of a more complete amino acid suite would imply $NH_3$ concentrations above a few weight percent, and (3) detection of sulfur-bearing amino acids (cysteine and methionine) would demonstrate additional sulfur cycling pathways in Titan's environment  (e.g., via atmospheric deposition or melt-rock interaction), which remains an open question (Nixon et al., 2025). Considering this, if we were to recommend a priority list for pre-flight DraMS testing with amino acid standards, we would recommend proline, alanine, β-alanine, cysteine, and methionine be included. Even though they are isomers, alanine and β-alanine can be separated by about a minute in retention time in the GC column with an inlet temperature increase of 10 °C/min (Pearce, Hörst, Sebree, et al., 2024). While equilibrium appears attainable within the deeper melt, the upper tens of meters accessible to *Dragonfly* likely experienced faster cooling and partial equilibration (see Section 2.4). Understanding how amino acids distribute across these freeze layers will also be important for interpreting surface detections but remains outside the scope of this study.

For detection strategy, DraMS can utilize LDMS and GC-MS in a complementary manner. LDMS is well-suited as a rapid screening tool: positive-ion mode enhances aromatic and basic amino acids (e.g., phenylalanine, tyrosine, tryptophan, histidine), while negative-ion mode favors acidic (e.g., aspartic acid, glutamic acid) and amide-bearing species (e.g., asparagine, glutamine). High-yield aliphatic products predicted by our models (alanine, glycine, β-alanine) should also be readily detectable in LDMS spectra. GC-MS provides added value where LDMS signals are weak or ambiguous: (i) enabling chiral separations (i.e., resolving enantiomers, e.g., D- vs. L-alanine); (ii) confirming low-volatility compounds via derivatization; and (iii) resolving isomeric interferences (e.g., distinguishing structural isomers like β-alanine from α-alanine).

Operationally, the finite number of single-use LDMS cups motivates their use in targeted, hypothesis-driven analyses rather than routine triage. The reusability of GC-MS cups may allow it to serve as the primary triage tool for high-throughput volatile screening, with LDMS reserved for especially promising or geologically significant samples. By aligning DraMS' sampling plan with Selk-specific thermodynamic predictions, *Dragonfly* can maximize science return: directly testing whether Titan's impact-induced aqueous environments can generate amino acids.

While this study is tailored to *Dragonfly's* Selk crater analyses, its thermodynamic framework naturally extends to other mission concepts. The POSEIDON concept couples a polar orbiter with a lake-lander and heavy drone, explicitly targeting Titan's high-latitude seas to probe HCN-driven prebiotic chemistry (Rodriguez et al., 2022) which complements *Dragonfly's* equatorial focus. Similarly, (G. Tobie et al., 2014) identifies impact-generated aqueous niches and inventories of complex organics as key objectives across icy moons like Titan and Enceladus, aligning closely with our work. Lastly, (Sulaiman et al., 2022) positions Titan (and Enceladus) chemistry as a strategic priority and highlights the need for quantitative baselines like those provided in our work. Although absolute yields will differ with local conditions, our models can be refined (e.g., adjusting initial concentrations and/or expected products) to guide target molecule selection and instrument preparation for future concepts and missions.

### 3.4. Assumptions and Limitations

We minimize total G at fixed T, P, and bulk elemental abundances using NASA-9 fits. The following assumptions are inherent to our analysis: (i) ideal solution baseline with unit activity coefficients (i.e., non-ideal ionic strength effects are neglected); (ii) homogeneous aqueous phase (without volatile loss or phase separation); (iii) no kinetic barriers; (iv) closed elemental system; (v) limited starting inventory (Table 4). We neglect photolysis, radiolysis, mineral catalysis, polymerization, adsorption, and oxidative sinks during the aqueous interval. Acid–base speciation (e.g., protonation states, zwitterionic forms) is not explicitly tracked; ammonia controls alkalinity but pH is not explicitly tracked or varied. We treat freezing point depression qualitatively and do not solve full phase equilibria (ice–brine partitioning). We assume freshwater melt pools (no added salts or minerals); however, impact melts on Titan may entrain salts and mineral fragments from its crust. Incorporating such solutes could: (i) depress the freezing point and thus prolong liquid lifetimes, (ii) buffer pH, (iii) alter amino acid speciation through metal-ion complexation, and/or (iv) catalyze reactions. These simplifications are appropriate for probing thermodynamic ceilings but may differ from melt conditions on Titan.

4. **Conclusion**

We find that Titan's post-impact melt pools could host amino acid synthesis. With $\geq 1\%$ $NH_3$ relative to water, nearly the full suite of canonical amino acids becomes thermodynamically accessible, whereas $NH_3$-free systems yield only alanine, β-alanine, and proline. Unlike the classic Strecker or aminonitrile hydrolysis pathways, our models reveal an energetically favorable route to alanine requiring only HCN, $C_2H_2$, and $H_2O$. In exploring this chemistry, we identify acrylonitrile as a plausible intermediate that can thermodynamically convert to alanine under aqueous conditions. These findings highlight both acetylene, which is rare on Earth but abundant on Titan, and acrylonitrile, as potential feedstocks and intermediates in prebiotic chemistry on Titan.

We emphasize that such modeling presents a framework to theoretically test new chemistries, before seeking out more resource and time-intensive experiments, especially for Titan conditions where cryoreactions may take years to reach equilibrium or reactants may be hard to work with in the lab. Nonetheless, future work may test such predictions experimentally and in situ. Laboratory studies simulating Titan-like aqueous environments with hydrogen cyanide and acetylene may help clarify the detailed intermediates involved in alanine formation and benchmark thermodynamic predictions against kinetic outcomes. Future kinetic studies should also consider the effects of temperature. Expanding our models to include sulfur chemistry, lipid and nucleobase synthesis, and time-dependent evolution will further constrain the range of molecular products expected in impact-induced habitats over Titan's history with the latter project providing a more concrete overview of yields at specific time steps in Titan's history. *Dragonfly's* DraMS instrument offers an opportunity to probe these questions directly: by targeting Selk crater, DraMS can evaluate whether Titan's surface supports amino acid synthesis, providing the first in situ test.

More broadly, this work demonstrates the importance of expanding our conception of prebiotic chemistry beyond Earth-centric feedstocks and pathways. Life's building blocks may arise not only from ammonia-rich Strecker networks or hydrothermal vent catalysis, but also from hydrocarbon-dominated chemistries unique to other worlds. Titan offers a natural laboratory where such alternative routes can be tested. However, the implications extend further: if amino acids can emerge from feedstocks like acetylene on Titan, then diverse chemistries across icy moons, asteroids, and even exoplanets may converge toward life's molecular precursors in a way dissimilar to early Earth. Identifying amino acids in Titan's crater environments would not only illuminate its chemical potential, but also broaden the search for life beyond Earth, reminding us that the pathways to biology may be as varied as the worlds that host them.

**Acknowledgements**

Our research relied on computational resources provided by the Negishi cluster, operated by the Rosen Center for Advanced Computing at Purdue University. We acknowledge the Purdue Community Cluster Program as described in (McCartney et al., 2014). We thank Dr. Kendra

Farnsworth for insightful discussion on the methodology in (Farnsworth et al., 2024), and Dr. Grayson Boyer for guidance on pyCHNOSZ. We also thank the two anonymous reviewers for their constructive feedback, which improved the quality of this manuscript.

**Financial Support**

I.M. was supported by the Frederick N. Andrews Fellowship awarded by Purdue University (2024-26).

**Author Contributions**

Ishaan Madan: Conceptualization, Data Curation, Formal Analysis, Investigation, Methodology, Project Administration, Resources, Software, Validation, Visualization, Writing – original draft, Writing – review & editing. Ben Pearce: Conceptualization, Methodology, Resources, Supervision, Validation, Writing – original draft, Writing – review & editing.

**Conflicts of Interest**

None

**Data Availability Statement**

All raw computational output files, data, and Python codes with the necessary data to reproduce any of the calculations in this work are openly available on Zenodo: https://doi.org/10.5281/zenodo.16921210.

**Appendix**

**Section A. Cantera Models Setup and Validation**
*Cantera Setup*

      In the VCS framework, a set of $C$ component species is first identified such that all other $N - C$ non-component species can be expressed as formal formation reactions from the components. The stoichiometric relationships are expressed through the $M$ x $N$ element matrix $A$, where $M$ is the number of conserved elements (including charge). The conservation constraints are:

$$A\,n = b \tag{A1}$$

where $n$ is the vector of species mole numbers and $b$ the total elemental abundances. Each independent reaction vector $v_j$ satisfies:

$$A\,v_j = 0 \tag{A2}$$

The number of independent reactions is $R = N - C$. For each independent reaction $j$, the reaction Gibbs free energy $(\Delta_r G_j)$ is:

$$\Delta_r G_j = \sum_{i=1}^{N} v_{i_j} \mu_i \tag{A3}$$

where $v_{i_j}$ is the stoichiometric coefficient of species $i$ in reaction $j$ (positive for products, negative for reactants) and $\mu_i$ is the chemical potential of species $i$, defined as:

$$\mu_i(T) = \mu_i^o(T) + RT\,ln\,a_i \tag{A4}$$

Here, $R$ is the gas constant, $T$ is the temperature and $a_i$ is the activity of species $i$. For a pure substance in its standard state, the standard-state chemical potential $\mu_i^o(T)$ is equal to the molar Gibbs free energy (shown below in Eq. A6). This follows directly from the definition of chemical potential $(\mu_i)$ as the partial molar Gibbs free energy:

$$\mu_i = \left(\frac{\partial G}{\partial n_i}\right)_{T,P,n_{j \neq i}} \tag{A5}$$

In a pure standard-state phase, adding 1 mole of $i$ increases the total Gibbs free energy by its molar Gibbs energy; hence:

$$\mu_i^o(T) = G_i^o(T) \tag{A6}$$

The standard Gibbs free energies of formation, $\Delta_f G^o(T)$ are obtained from the CHNOSZ database (Dick, 2019) via the python wrapper, pyCHNOSZ (Boyer, 2024). For molecules without published Gibbs data, we estimate $\Delta_f G^o(T)$ using the Gibbs free energies estimator methods described in Section 2.2 of the manuscript. These $\Delta_f G^o(T)$ values are fit to the NASA 9-coefficient polynomial format (McBride et al., 2002), enabling Cantera to reconstruct $\mu_i^o(T)$ at any temperature during the calculation. With $\mu_i^o(T)$ available for all species, Cantera evaluates each $\mu_i(T)$ via Eq. A4 and computes $\Delta_r G_j$ for all independent reactions using Eq. A3. The solver then adjusts the mole numbers $n_i$ to minimize the total Gibbs energy:

$$G(n) = \sum_{i=1}^{N} n_i \mu_i \tag{A7}$$

In practice, the minimization is performed under the constraints that (i) all elemental abundances are conserved (Eq. A1) and (ii) all mole numbers remain non-negative ($n_i \geq 0$ for all $i$).

*Polynomial Fitting*

To interface with Cantera's solver as described, we convert $\Delta_f G^o(T)$ data into the NASA 9-coefficient polynomial format (McBride et al., 2002), which defines the heat capacity, enthalpy, and entropy over a specified temperature range according to:

$$\frac{c_p^o(T)}{R} = a_0 T^{-2} + a_1 T^{-1} + a_2 + a_3 T + a_4 T^2 + a_5 T^3 + a_6 T^4 \tag{A8}$$

$$\frac{h^o(T)}{RT} = -a_0 T^{-2} + a_1 \frac{ln\,T}{T} + a_2 + \frac{a_3}{2}T + \frac{a_4}{3}T^2 + \frac{a_5}{4}T^3 + \frac{a_6}{5}T^4 + \frac{a_7}{T} \tag{A9}$$

$$\frac{s^o(T)}{R} = -\frac{a_0}{2}T^{-2} - a_1 T^{-1} + a_2 ln\,T + a_3 T + \frac{a_4}{2}T^2 + \frac{a_5}{3}T^3 + \frac{a_6}{4}T^4 + a_8 \tag{A10}$$

Equations A8-10 capture heat capacity ($c_p^o(T)$), enthalpy ($h^o(T)$), and entropy ($s^o(T)$) simultaneously. However, directly fitting all three functions is challenging, especially when working from $\Delta_f G^o(T)$ data rather than experimental $c_p$ measurements. To address this, we use the thermodynamic identity:

$$G(T) = H(T) - TS(T) \tag{A11}$$

Where $H(T)$ is the enthalpy, $T$ is the temperature, and $S(T)$ is the entropy. We algebraically combine Eq. A9 and 10 via Eq. A11 to obtain a single expression for $\Delta_f G^o(T)$ in terms of the coefficients $a_0 - a_8$, then perform a single least-squares fit to the $\Delta_f G^o(T)$ curve from CHNOSZ or our quantum-chemistry estimates. While Eq. A8 for $c_p^o(T)$ is not explicitly used in our fitting, the coefficients obtained from the $\Delta_f G^o(T)$ fit are internally used by Cantera to compute $c_p^o(T)$ internally. This avoids instabilities inherent to multi-objective fitting and ensures direct consistency between the input data and Cantera's minimization framework. The resulting NASA-9 coefficients are encoded into YAML species files compatible with Cantera's equilibrium solver.

*Validation*

Validation of the Cantera framework against the thermochemical study of (Pearce & Pudritz, 2016) is carried out under identical conditions. Pearce and Pudritz (2016) modeled nucleobase synthesis in planetesimal interiors using Gibbs free energies from an earlier version of the CHNOSZ database and a proprietary thermochemical code called ChemApp. We fix the pressure at 100 bar, matching their static pressure for planetesimal interiors, and we initialize each simulation with the mole fractions given in their Table 3. Our Gibbs free energies are obtained from the most recent CHNOSZ database via pyCHNOSZ (Boyer, 2024; Shock & Boyer, 2020), and fit to the NASA-9 polynomials. Across the temperature window of 0–250 ºC, the mass yields predicted by Cantera (olive circles in Fig. A1) reproduce Pearce and Pudritz values extremely closely for all four reactions, with the maximum deviation being <5% across all reactions. The max errors are: 4.5% (5 HCN → adenine), 1.2% (HCN + NH$_3$ + H$_2$O → adenine), 0.0% (3 HCN + CH$_2$O → cytosine), and 4.4% (5 HCN + H$_2$O → guanine + H$_2$). This excellent agreement corroborates both: our NASA-9 polynomial fits derived from the CHNOSZ dataset and Cantera's VCS minimization scheme, and it establishes that, under liquid-water conditions, the two approaches return virtually indistinguishable equilibria.

Figure A1. Thermodynamic validation of our Cantera framework against Pearce and Pudritz (2016). Equilibrium yields (expressed as kg product/kg water) are plotted for four nucleobase-forming reactions: (a) 5 HCN → adenine, (b) HCN + NH$_3$ + H$_2$O → adenine, (c) 3 HCN + CH$_2$O → cytosine, and (d) 5 HCN + H$_2$O → guanine + H$_2$. Olive solid lines with circle markers show the results of our Cantera equilibrium calculations using NASA-9 polynomial coefficients derived from CHNOSZ data, while black dashed lines with square markers reproduce the mass-yield trends reported by Pearce and Pudritz (2016). The maximum deviation across all reactions within 0–250 ºC is <5%. The vertical dashed grey line at 311 ºC denotes the boiling point of water at a pressure of 100 bar.

Prebiotic Chemistry in Selk Crater on Titan

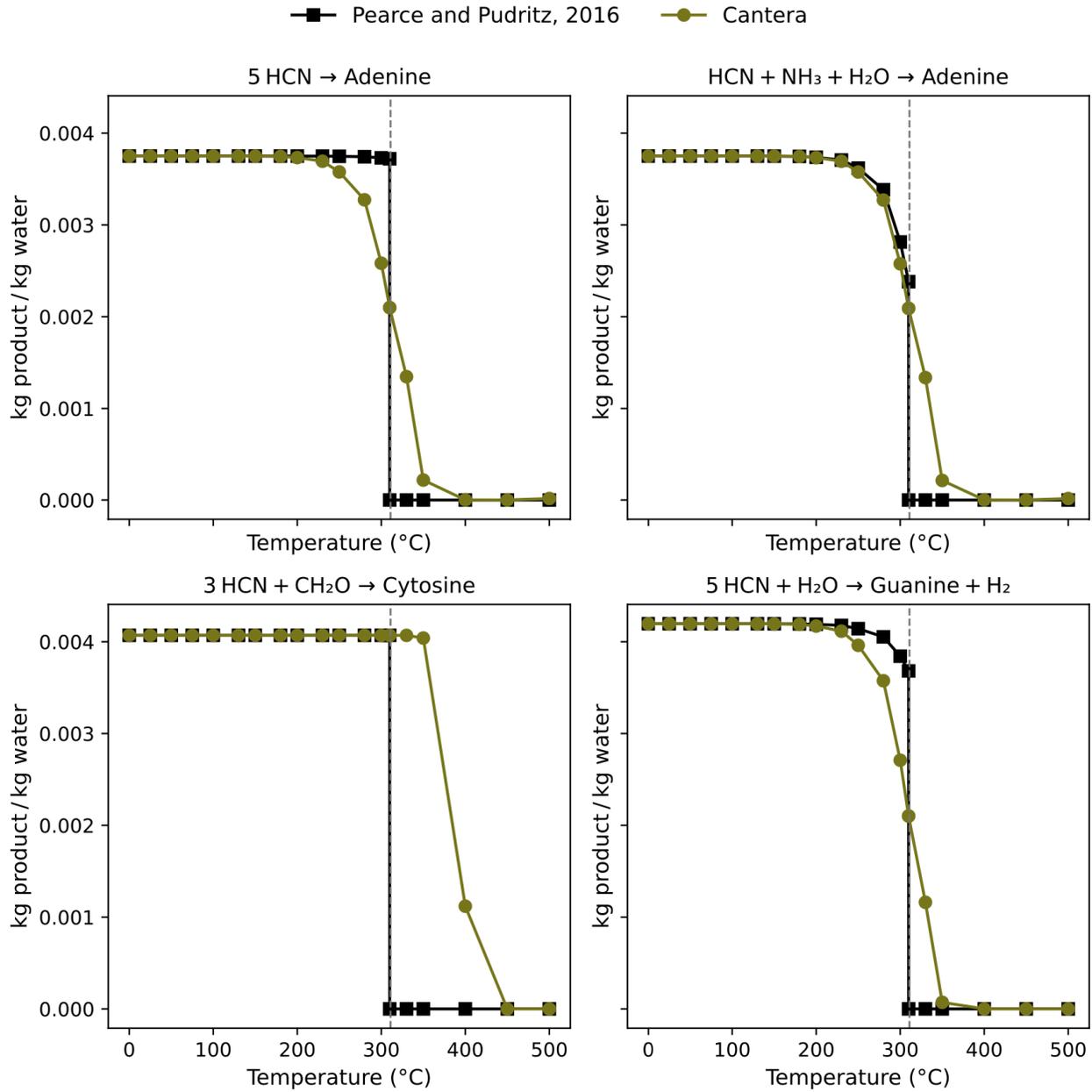

Deviations emerge in the vicinity of the liquid-vapor transition (boiling point) of water at ~311 ºC for 100 bar. Whereas Pearce and Pudritz (2016) report a sharp, discontinuous drop in product abundance immediately past this boiling point, our Cantera calculations decline more gradually. The most likely explanation is updates to the CHNOSZ database since 2014, particularly in the values of Gibbs free energies at elevated temperatures and pressures. This is supported by a direct comparison of raw data (specifically for adenine) between the 2014 and 2025 CHNOSZ versions, which shows differences in the Gibbs energies near the boiling point. These differences may be amplified by the way each model handles thermodynamic discontinuities in the near-critical region, where shifts in solvent properties may become abrupt.

Ultimately, for the purposes of our Titan impact-pond application, these high-temperature discrepancies are inconsequential. The thermodynamic equilibrium we seek to model occurs well below 100 ºC, a range in which the Cantera predictions are indistinguishable from the validated Pearce and Pudritz (2016) results. We therefore conclude that: (i) our NASA-9 coefficients encode valid and robust aqueous thermodynamic data, and (ii) Cantera provides a reliable thermodynamic framework for exploring organic synthesis under the aqueous post-impact conditions of icy moons like Titan.

**Section B. Gibbs Free Energies Estimator**
To estimate Gibbs free energies of formation for molecules lacking published thermodynamic data, we apply the three step protocol of (Ochterski, 2000) at 298 K:

$$\Delta_f H^o(M, 0K) = \sum_{atoms} x \Delta_f H^o(X, 0K) - \sum_{atoms} x E_0(X) - E_0(M) + E_{ZPE}(M) \qquad (A12)$$

$$\Delta_f H^o(M, 298K) = \Delta_f H^o(M, 0K) + (H_M^o(298K) - H_M^o(0K)) - \sum_{atoms} x(H_X^o(298K) - H_X^o(0K)) \qquad (A13)$$

$$\Delta_f G^o(M, 298K) = \Delta_f H^o(298K) - T(S^o(M, 298K) - \Sigma S^o(X, 298K)) \qquad (A14)$$

Where $\Delta_f H^o$ is the enthalpy of formation, $E_0$ is the electronic energy, $E_{ZPE}$ is the zero-point energy correction, $H^o(T)$ is the thermal enthalpy correction at temperature $T$, $\Delta_f G^o$ is the Gibbs free energy of formation, $T$ is the temperature (here, 298 K), $S^o$ is the entropy (sum of translational, rotational, vibrational contributions), $M$ is the target molecule being formed, $X$ is a reference atom (e.g. C, H, O, N), and $x$ is the number of atoms of type $X$. Please refer to (Ochterski, 2000) for any further details.

All quantum chemical (QC) calculations methods are described in the main manuscript. We perform full calculations for C, $H_2$, $O_2$, $N_2$, glycine, and alanine across the temperature window 273–598 K. Although only the 298 K values are ultimately used in our estimator, these extended calculations are shown here to highlight a key issue: the direct QC trends across temperature exhibit a monotonic increase in $\Delta_f G°$ that is inconsistent with the CHNOSZ database. This discrepancy is illustrated for glycine in Figure A2. For aminoacetonitrile (AAH), 2-aminopropanenitrile (2-AH), β-alanine, and acrylonitrile, calculations are performed only at 298 K. Their temperature-dependent Gibbs free energies are then derived using the analog slope method described in the main manuscript. The resulting values are reported in Table A1.

Figure A2. Validation of the Gibbs free energy estimator using glycine. Black line shows the raw quantum chemical results across temperature, which exhibit a monotonic increase. The red line applies the analog-slope correction described in the main manuscript, which matches the CHNOSZ trend to within 17%. Olive square markers show the reference CHNOSZ database values.

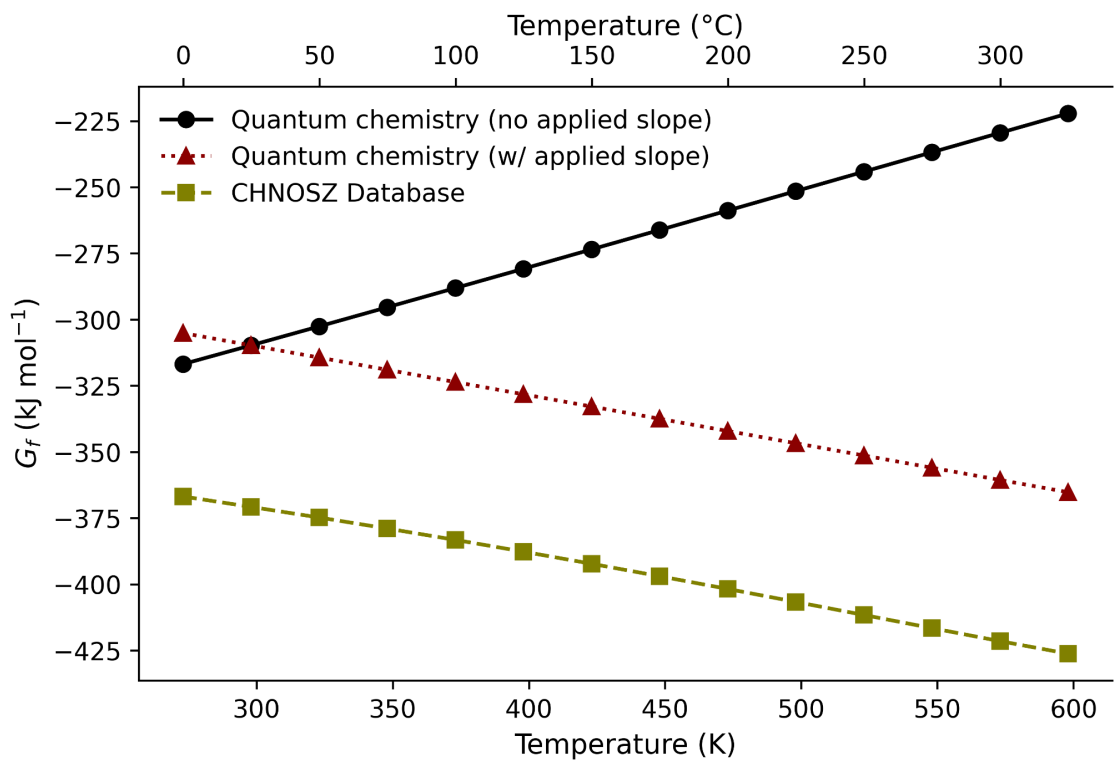

Table A1. Calculated standard Gibbs free energies of formation, $\Delta_f G^o$ (J mol$^{-1}$) for aminoacetonitrile (AAH), 2-aminopropanenitrile (2AH), β-alanine, and acrylonitrile as a function of temperature.

| | $\Delta_f G^o$ (J mol$^{-1}$) | | | |
|---|---|---|---|---|
| T (°C) | AAH | 2AH | β-alanine | Acrylonitrile |
| 0.01 | 141727 | 154434 | -285049 | 179558 |
| 10 | 139879 | 152206 | -287277 | 177330 |
| 20 | 138029 | 149976 | -289507 | 175100 |
| 30 | 136179 | 147746 | -291737 | 172870 |
| 40 | 134329 | 145516 | -293967 | 170640 |
| 50 | 132479 | 143286 | -296197 | 168410 |
| 60 | 130629 | 141056 | -298427 | 166180 |
| 70 | 128779 | 138826 | -300658 | 163950 |
| 80 | 126929 | 136596 | -302888 | 161720 |
| 90 | 125079 | 134366 | -305118 | 159490 |
| 100 | 123229 | 132136 | -307348 | 157260 |
| 110 | 121379 | 129906 | -309578 | 155030 |
| 120 | 119529 | 127676 | -311808 | 152800 |
| 130 | 117679 | 125446 | -314038 | 150570 |
| 140 | 115829 | 123216 | -316268 | 148340 |
| 150 | 113979 | 120986 | -318498 | 146110 |
| 160 | 112129 | 118756 | -320728 | 143880 |
| 170 | 110279 | 116526 | -322958 | 141650 |
| 180 | 108429 | 114296 | -325188 | 139420 |
| 190 | 106579 | 112066 | -327418 | 137190 |
| 200 | 104729 | 109836 | -329648 | 134960 |
| 210 | 102879 | 107606 | -331879 | 132730 |
| 220 | 101029 | 105376 | -334109 | 130500 |
| 230 | 99179 | 103146 | -336339 | 128270 |
| 240 | 97329 | 100916 | -338569 | 126040 |
| 250 | 95479 | 98686 | -340799 | 123810 |
| 260 | 93629 | 96456 | -343029 | 121580 |
| 270 | 91779 | 94226 | -345259 | 119350 |
| 280 | 89929 | 91996 | -347489 | 117120 |
| 290 | 88079 | 89766 | -349719 | 114890 |
| 300 | 86229 | 87536 | -351949 | 112660 |
| 310 | 84379 | 85306 | -354179 | 110430 |
| 320 | 82529 | 83076 | -356409 | 108200 |
| 330 | 80679 | 80846 | -358639 | 105970 |
| 340 | 78829 | 78616 | -360869 | 103740 |
| 350 | 76979 | 76386 | -363100 | 101510 |
| 360 | 75129 | 74156 | -365330 | 99280 |
| 370 | 73279 | 71926 | -367560 | 97050 |

## Section C. Why Thermodynamics for Titan?

*Equilibration Time Calculations*

Equilibration times for glycine and alanine are calculated by integrating the standard first-order rate law, defining the converted fraction, and solving for time:

$$[A]_t = [A]_0 e^{-kt} \Rightarrow f = \frac{[A]_0 - [A]_t}{[A]_0} = 1 - e^{-kt} \Rightarrow t = \frac{-ln(1-f)}{k}$$

where $[A]_t$ is the concentration of species $A$ at time $t$, $[A]_0$ is the initial concentration, $k$ is the first-order rate constant (time$^{-1}$), $t$ is the time elapsed, $f$ is the fraction of $A$ that has converted.

## Section D. Applying the model to Titan: Melt Volume and Input Organics

We estimate the total starting inventories of HCN, $C_2H_2$, $NH_3$, and $H_2O$ within Selk crater's melt pool. Surface flux values for $C_2H_2$ and HCN from (Neish et al., 2024) are converted to total moles using the 45-km radius catchment area, an accumulation time of 4.3 Gyr, and each species' molar mass (as described in the manuscript). A 10% survival factor is then applied, and $NH_3$ is included parametrically relative to the total water inventory. The resulting parameters and normalized concentrations are summarized in Table A2.

Table A2. Summary of parameters used to derive the initial reactant concentrations for Selk crater melt pool models.

| Species | Flux (g cm$^{-2}$ Gyr$^{-1}$)[1] | Molar Mass (g mol$^{-1}$) | Final Moles (mol) | Normalized (mol/mol $H_2O$) |
|---|---|---|---|---|
| $C_2H_2$ | 439 | 26.04 | $4.61 \times 10^{14}$ | 0.042 |
| HCN | 218 | 27.03 | $2.21 \times 10^{14}$ | 0.020 |
| $NH_3$ | - | 17.03 | - | 0, 0.01, 0.02, 0.03, 0.04, 0.05, 0.10 |
| $H_2O$ | - | 18.02 | $1.11 \times 10^{16}$ | 1.00 |

[1](Neish et al., 2024).

## Section E: Survival Factor Sensitivity Tests

To evaluate how uncertainties in the initial organic inventory affect amino acid yields, we perform sensitivity tests varying the survival factor. While our fiducial models adopt a 10% survival factor, here we explore two scenarios with survival factors of 1% and 30%. The sensitivity tests show broadly consistent results across most amino acids and ammonia concentrations (Figure A3, Tables A3 and A4). For the 1% survival factor, yields are nearly identical to the fiducial case, with modest yield increases for isoleucine, leucine, and valine with 1% $NH_3$. In contrast, the 30% survival factor runs produced slightly lower yields for the same species, as well as for arginine and lysine within the 1–3% NH$_3$ range. At higher ammonia concentrations ($\geq$ 5%), yields among all species converge toward the fiducial case. These tests assess how equilibrium yields scale with the total available reactant inventory and confirm that overall, the pattern of amino acid accessibility ($\geq$ 1% NH$_3$) and yield tapering beyond 2% NH$_3$ remains unchanged.

Figure A3. Heat maps of amino acid percent yields at 0 ºC relative to the starting HCN inventory with 1% and 30% survival factors of initial organic inventory. Columns correspond to varying NH$_3$ concentrations (0–10% of the total water content). Rows list the 21 amino acids evaluated. Cell shading follows a viridis scale from 0% (purple) to 100% (yellow).

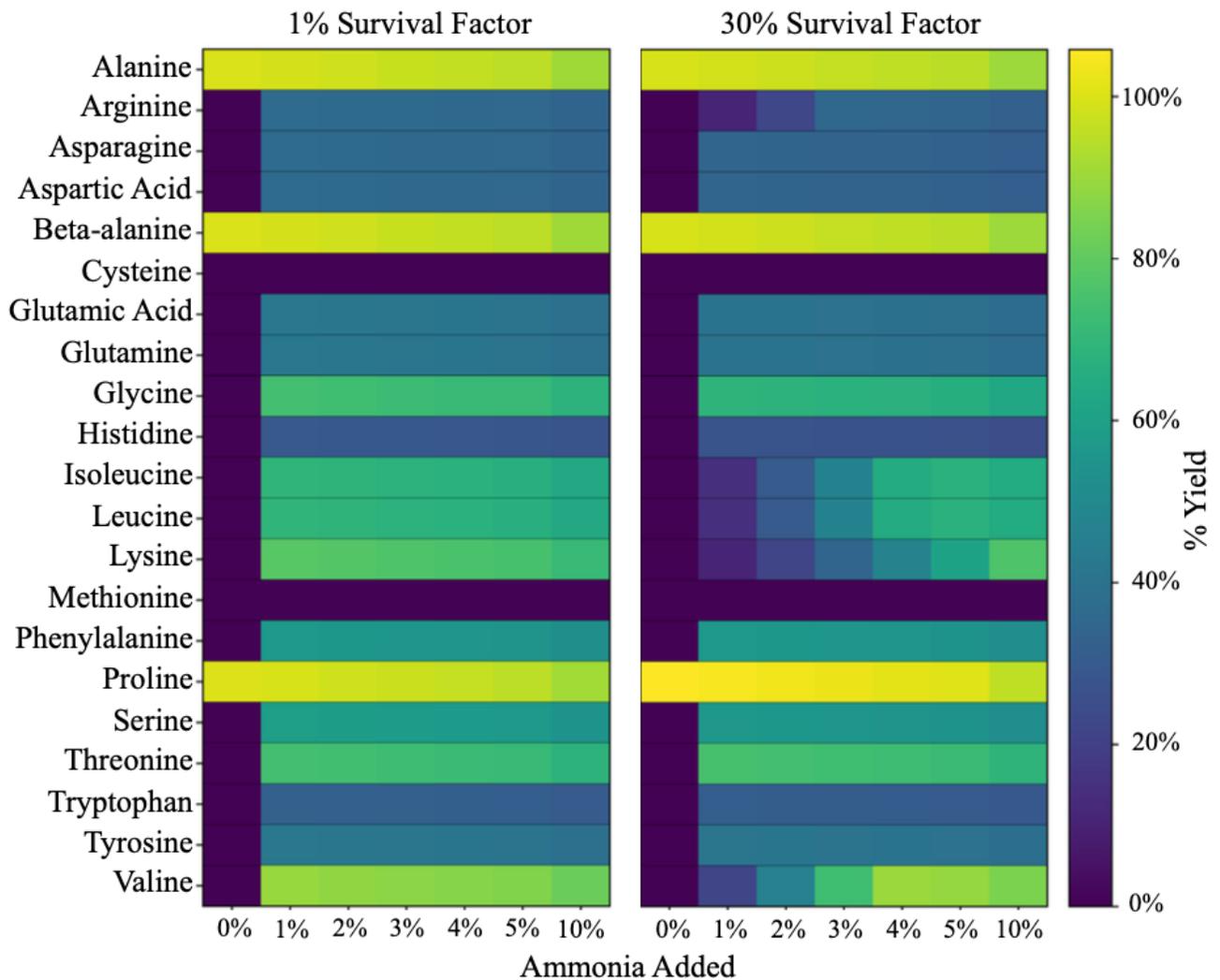

Table A3. Percent yields of amino acids at 0 ºC, expressed relative to the initial HCN inventory where the survival factor of the starting inventory is 1%.

| Amino Acid | % Yield (with respect to HCN) | | | | | | |
|---|---|---|---|---|---|---|---|
| | *0% NH₃* | *1% NH₃* | *2% NH₃* | *3% NH₃* | *4% NH₃* | *5% NH₃* | *10% NH₃* |
| Alanine | 100.0 | 99.0 | 98.0 | 97.1 | 96.1 | 95.2 | 90.9 |
| Arginine | 0.0 | 37.1 | 36.7 | 36.4 | 36.0 | 35.7 | 34.1 |
| Asparagine | 0.0 | 37.0 | 36.7 | 36.3 | 36.0 | 35.6 | 34.0 |
| Aspartic acid | 0.0 | 37.0 | 36.7 | 36.3 | 36.0 | 35.6 | 34.0 |
| β-alanine | 100.0 | 99.0 | 98.0 | 97.1 | 96.1 | 95.2 | 90.9 |
| Cysteine | 0.0 | 0.0 | 0.0 | 0.0 | 0.0 | 0.0 | 0.0 |
| Glutamic acid | 0.0 | 42.4 | 41.9 | 41.5 | 41.1 | 40.8 | 38.9 |
| Glutamine | 0.0 | 42.4 | 41.9 | 41.5 | 41.1 | 40.8 | 38.9 |
| Glycine | 0.0 | 74.1 | 73.3 | 72.6 | 71.9 | 71.2 | 68.0 |
| Histidine | 0.0 | 29.6 | 29.3 | 29.0 | 28.8 | 28.5 | 27.2 |
| Isoleucine | 0.0 | 69.4 | 68.7 | 68.0 | 67.4 | 66.7 | 63.7 |
| Leucine | 0.0 | 69.4 | 68.7 | 68.0 | 67.4 | 66.7 | 63.7 |
| Lysine | 0.0 | 78.2 | 77.4 | 76.7 | 75.9 | 75.2 | 71.8 |
| Methionine | 0.0 | 0.0 | 0.0 | 0.0 | 0.0 | 0.0 | 0.0 |
| Phenylalanine | 0.0 | 56.7 | 56.2 | 55.6 | 55.1 | 54.6 | 52.1 |
| Proline | 100.2 | 99.2 | 98.2 | 97.3 | 96.3 | 95.4 | 91.1 |
| Serine | 0.0 | 59.3 | 58.7 | 58.2 | 57.6 | 57.0 | 54.5 |
| Threonine | 0.0 | 74.3 | 73.6 | 72.8 | 72.1 | 71.4 | 68.2 |
| Tryptophan | 0.0 | 33.0 | 32.6 | 32.3 | 32.0 | 31.7 | 30.3 |
| Tyrosine | 0.0 | 42.4 | 42.0 | 41.6 | 41.2 | 40.8 | 38.9 |
| Valine | 0.0 | 89.3 | 88.4 | 87.5 | 86.7 | 85.8 | 81.9 |

Table A4. Percent yields of amino acids at 0 ºC, expressed relative to the initial HCN inventory where the survival factor of the starting inventory is 30%.

| Amino Acid | % Yield (with respect to HCN) | | | | | | |
|---|---|---|---|---|---|---|---|
| | *0% NH₃* | *1% NH₃* | *2% NH₃* | *3% NH₃* | *4% NH₃* | *5% NH₃* | *10% NH₃* |
| Alanine | 99.5 | 98.5 | 97.6 | 96.6 | 95.7 | 94.8 | 90.5 |
| Arginine | 0.0 | 10.9 | 22.6 | 35.2 | 34.9 | 34.6 | 33.0 |
| Asparagine | 0.0 | 34.4 | 34.1 | 33.8 | 33.5 | 33.2 | 31.8 |
| Aspartic acid | 0.0 | 34.4 | 34.1 | 33.8 | 33.5 | 33.2 | 31.8 |
| β-alanine | 99.5 | 98.5 | 97.6 | 96.6 | 95.7 | 94.8 | 90.5 |
| Cysteine | 0.0 | 0.0 | 0.0 | 0.0 | 0.0 | 0.0 | 0.0 |
| Glutamic acid | 0.0 | 40.5 | 40.1 | 39.8 | 39.4 | 39.0 | 37.3 |
| Glutamine | 0.0 | 40.5 | 40.1 | 39.8 | 39.4 | 39.0 | 37.3 |
| Glycine | 0.0 | 68.8 | 68.2 | 67.6 | 67.0 | 66.4 | 63.6 |
| Histidine | 0.0 | 27.3 | 27.1 | 26.8 | 26.6 | 26.3 | 25.2 |
| Isoleucine | 0.0 | 14.6 | 30.2 | 46.9 | 65.0 | 67.6 | 64.5 |
| Leucine | 0.0 | 14.6 | 30.2 | 46.9 | 65.0 | 67.6 | 64.5 |
| Lysine | 0.0 | 10.8 | 22.3 | 34.5 | 47.4 | 61.1 | 76.6 |
| Methionine | 0.0 | 0.0 | 0.0 | 0.0 | 0.0 | 0.0 | 0.0 |
| Phenylalanine | 0.0 | 56.7 | 56.2 | 55.6 | 55.1 | 54.5 | 52.0 |
| Proline | 105.8 | 104.7 | 103.6 | 102.6 | 101.5 | 100.5 | 95.7 |
| Serine | 0.0 | 56.4 | 55.9 | 55.4 | 54.9 | 54.4 | 52.0 |
| Threonine | 0.0 | 75.0 | 74.3 | 73.5 | 72.8 | 72.1 | 68.8 |
| Tryptophan | 0.0 | 31.6 | 31.3 | 31.0 | 30.7 | 30.4 | 29.1 |
| Tyrosine | 0.0 | 41.5 | 41.1 | 40.7 | 40.3 | 40.0 | 38.2 |
| Valine | 0.0 | 22.1 | 46.5 | 73.5 | 89.9 | 89.0 | 84.8 |

**Appendix References**